%% file: main.tex
\documentclass[times, twoside]{StyleArXiv}
\usepackage{booktabs}
\usepackage{multicol,multirow}
\usepackage{orcidlink}
\usepackage{subcaption}
\usepackage{cleveref}
\usepackage{graphicx}%
\usepackage{multirow}%
\usepackage{amsmath,amssymb,amsfonts}%

\leadauthor{Siddhad and Meena} 

\begin{document}

\title{SwitchBraidNet: Quantisation-Aware Lightweight Architecture for Hybrid Brain-Computer Interface}
\shorttitle{SwitchBraidNet: Quantisation-Aware Lightweight Architecture for Hybrid Brain-Computer Interface}

\author[1,\Letter]{Gourav~Siddhad~\orcidlink{0000-0001-5883-3863}}
\author[1]{Yogesh~Kumar~Meena~\orcidlink{0000-0003-3429-8426}}
\affil[1]{Human-AI Interaction (HAIx) Lab, Indian Institute of Technology Gandhinagar, Gujarat, 382055, India}

\maketitle



\begin{abstract}
Hybrid brain-computer interfaces (BCIs) that integrate motor imagery (MI) and steady-state visual evoked potentials (SSVEP) provide high-dimensional neural decoding but typically exceed the computational limits of embedded hardware. To address this, we propose SwitchBraidNet, a compact EEG classification architecture designed for low-power deployment. The model employs a dual-path temporal braid to extract multiscale oscillatory features, an adaptive squeeze-and-excitation spatial switch for electrode gating, and a log-variance readout layer for direct band-power encoding. Furthermore, through systematic quantisation-aware training on the OpenBMI dataset, we compared SwitchBraidNet against four established baselines across FP32, FP16, and INT8 precisions. Experimental results demonstrate superior efficiency and performance, achieving MI accuracy of 69.49\% (FP16), SSVEP accuracy of 93.48\% (FP32), and a hybrid information transfer rate of 64.82 bits/min (FP16). With an INT8 footprint of only 3.03 KB, SwitchBraidNet maintains high accuracy across varying numerical precisions, demonstrating its suitability for low-power embedded BCI deployment. 
\end{abstract}


\begin{corrauthor}
drgsiddhad{\at}gmail.com
\end{corrauthor}


\section{Introduction}
\label{sec_intro}

Brain--computer interfaces (BCIs) establish a direct communication pathway between the human nervous system and external devices, bypassing conventional neuromuscular output channels~\cite{wolpaw2000brain}. Among EEG-based paradigms, motor imagery (MI) and steady-state visual evoked potentials (SSVEP) are the two most clinically viable approaches. MI exploits sensorimotor $\mu$-band ($8$--$13$~Hz) and $\beta$-band ($13$--$30$~Hz) desynchronisation during imagined movements~\cite{pfurtscheller2001motor}, whereas SSVEP leverages phase-locked occipital neural responses driven by flickering visual stimuli~\cite{lee2019eeg}.

Both paradigms have been validated in assistive technology applications ranging from wheelchair control to text spellers and robotic rehabilitation devices~\cite{pfurtscheller2010hybrid, zhang2025hybrid}. Their combination has been explored across a range of hybrid architectures, including gaze-MI systems that decouple visual target selection from motor-triggered confirmation~\cite{o2014exploring}, to further extend the scope of non-muscular communication. Nevertheless, both paradigms carry individual limitations: MI offers flexible voluntary control but suffers from low signal-to-noise ratio and high inter-subject variability, with BCI inefficiency rates as high as 53.7\% reported on large-scale datasets~\cite{lee2019eeg}, whereas SSVEP achieves high information throughput but demands continuous fixation on flickering stimuli, which is cognitively fatiguing for extended use~\cite{luo2023hybrid}.

These complementary limitations have motivated the development of hybrid BCI (hBCI) systems that fuse signals from both paradigms to expand the effective command space and improve system robustness~\cite{pfurtscheller2010hybrid}. Crucially, users who cannot operate an ERD-based BCI can often control an SSVEP-based system and vice versa, substantially reducing overall user inefficiency~\cite{pfurtscheller2010hybrid, lee2019eeg}. Poor performance, often labelled as inefficiency, frequently stems from external factors like insufficient training, suboptimal calibration, or flawed experimental design, rather than inherent user inability. Recent work has extended this principle with peripheral visual field stimulation to improve user comfort while retaining SSVEP decoding accuracy~\cite{zhang2025hybrid}, and with multi-modal combinations of MI, SSVEP, and overt spatial attention (OSA) to further enrich the neural feature space~\cite{zhang2025novel}.

Despite the rapidly growing literature on deep learning for EEG-based BCIs, three challenges have not been addressed jointly. First, the neural signatures of MI and SSVEP occupy different spectral, spatial, and temporal subspaces, requiring feature representations that can accommodate heterogeneous signal statistics within a single model~\cite{luo2023hybrid, lee2019eeg}. Second, practical BCI deployment targets wearable or implantable hardware: microcontrollers, FPGAs, or neuromorphic chips, where memory, power, and compute budgets are severely constrained~\cite{cecotti2024post, meena2015towards}. Importantly, while \cite{meena2015towards} simultaneously records MI and gaze signals in a live hybrid paradigm, \cite{cecotti2024post} applies quantisation offline to a previously trained model, reflecting a divide in the literature between real-time simultaneous acquisition and post-hoc model compression. Third, quantising a trained model to low bit-widths, as required for embedded deployment, introduces quantisation noise that can disproportionately degrade architectures whose internal activations span a wide dynamic range~\cite{cecotti2024post}.

\begin{figure*}[!t]
    \centering
    \includegraphics[width=0.7\linewidth]{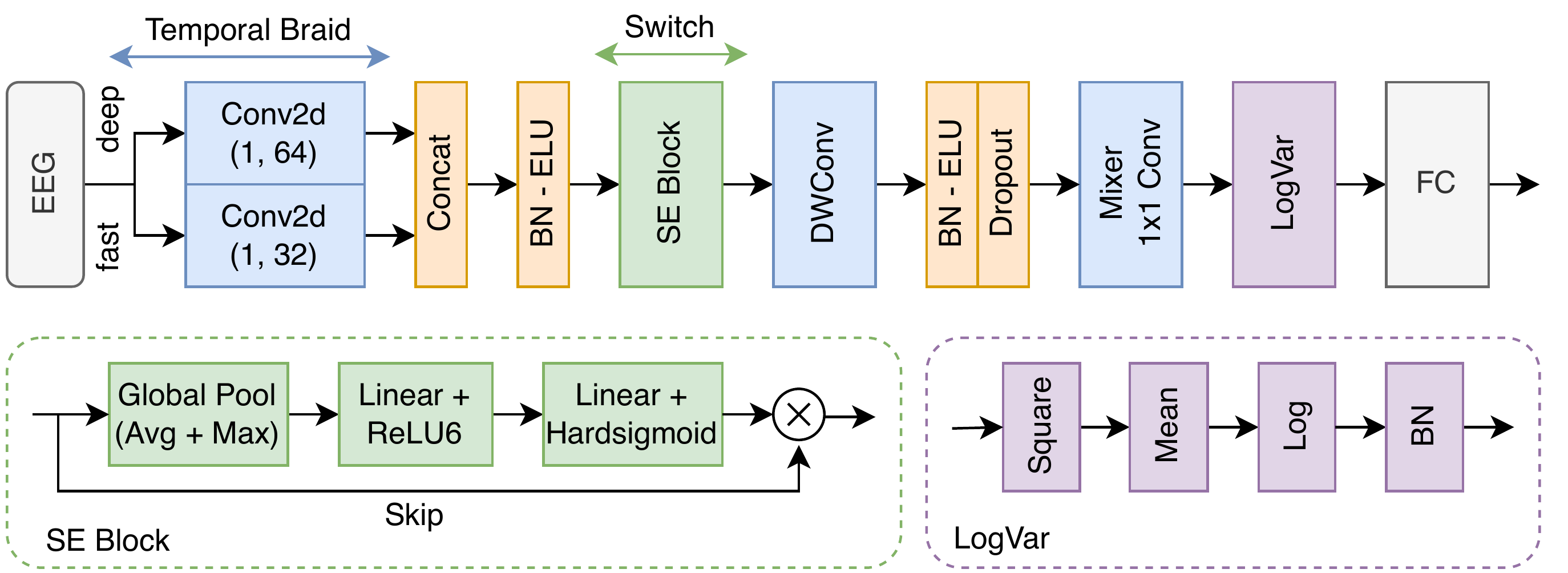}
    \caption{Proposed SwitchBraidNet Architecture. The model employs a dual-path temporal braid using deep and fast convolutional kernels to extract multi-scale neural features. The spatial switch utilises a squeeze-and-excitation (SE) block to dynamically weigh electrode importance, followed by a hardware-efficient LogVarLayer that collapses the temporal dimension into stable, low-bit power features suitable for 8-bit QAT optimisation.}
    \label{fig_arch}
    \vspace{-3mm}
\end{figure*}

Existing benchmarks typically evaluate full-precision (FP32) models on a single paradigm, and the few works that consider model compression focus on post-training quantisation (PTQ) rather than quantisation-aware training (QAT)~\cite{cecotti2024post}, which has been shown to yield significantly better accuracy retention at INT8 and below~\cite{jacob2018quantization}. Also, hBCI evaluations rarely formalise the relationship between classification accuracy and information transfer rate (ITR) across sequential vs simultaneous modes. The scoping literature on hBCIs confirms that simultaneous mode systems are the majority design choice, yet the ITR implications relative to sequential operation are not analytically derived or validated~\cite{mussi2022eeg}. To address these challenges, this work provides two major contributions, by:
\begin{itemize}
    \item Proposing SwitchBraidNet, a new lightweight EEG deep learning architecture for the MI-SSVEP hybrid brain-computer interface (hBCI), designed for physiological and hardware constraints with end-to-end quantisation robustness. 
    \item Presenting the first systematic quantisation-aware training benchmark on the OpenBMI dataset~\cite{lee2019eeg}, comparing five architectures at FP32, FP16, and INT8 across MI, SSVEP, and hybrid paradigms.
\end{itemize}


\section{Materials and Methods}
\label{sec_method}

This section describes the proposed architecture, hBCI framework, quantisation strategy, dataset, preprocessing, experimental setup, and evaluation metrics used in this study. To ensure transparency and reproducibility, source code is available at the SwitchBraidNet\footnote{https://github.com/HAIx-Lab/SwitchBraidNet} repository.

\subsection{SwitchBraidNet}

SwitchBraidNet is a lightweight architecture, designed from the ground up for hardware-aware deployment. It comprises three functional stages (Fig.~\ref{fig_arch}): a dual-path temporal braid, a spatial switch, and a log-variance readout head.

\subsubsection{Dual-Path Temporal Braid} SwitchBraidNet employs two parallel temporal convolution paths operating at distinct temporal scales. A deep path uses a $1 \times 64$ convolution (with padding 32) that captures slow oscillatory content (i.e., lower-frequency bands). A fast path uses a $1 \times 32$ convolution (with 16-pixel padding) tuned for faster transients. The outputs of both paths are concatenated along the channel dimension to produce a joint representation, followed by batch normalisation (BN) and ELU activation.

\subsubsection{Spatial Switch via Squeeze-and-Excitation} The spatial stage applies a squeeze-and-excitation (SE) block \cite{hu2018squeeze} prior to spatial convolution. This serves as a learned switch that gates electrode contributions by their global relevance. The SE block computes both average-pooled and max-pooled statistics across the spatial-temporal dimensions, combines them, and passes the result through a two-layer bottleneck (reduction ratio $r = 4$) with ReLU6 and hardsigmoid activations. The latter is advantageous for low-bit quantisation due to its piecewise-linear form. The SE output scales the feature maps via channel-wise multiplication before depthwise spatial convolution (kernel $C \times 1$, $\text{groups}=32$). This reduces the spatial dimension from $C$ to 1 while maintaining channel independence. Spatial BN, ELU activation, and 2-D dropout (0.2) follow. A $1 \times 1$ pointwise mixer convolution then projects from 32 to 16 channels with BN and ELU, acting as a learned feature recombination stage.

\subsubsection{Log-Variance Readout} SwitchBraidNet employs a LogVarLayer that collapses the temporal dimension by computing the mean squared amplitude across time-equivalent to a mean-power estimate, followed by a log transform with numerical stabilisation ($\epsilon = 10^{-6}$) and a BN layer:
\begin{equation}
\label{eq_logvar}
    v = \log( E_t[x^2] + \epsilon ), \quad \text{BN}(v)
\end{equation}

This design explicitly targets the power of neural oscillations, replicating the essential computation of logarithmic band-power features within a differentiable, quantisation-friendly layer. The BN within LogVarLayer stabilises the dynamic range of the output, which is critical for ensuring model accuracy when compressed to low-bit precision. The resulting 16-dimensional feature vector is passed to a linear classification head.

\subsection{Hybrid BCI Framework}

\begin{figure}[!t]
    \centering
    \begin{subfigure}[b]{0.56\linewidth}
        \centering
        \includegraphics[width=\linewidth]{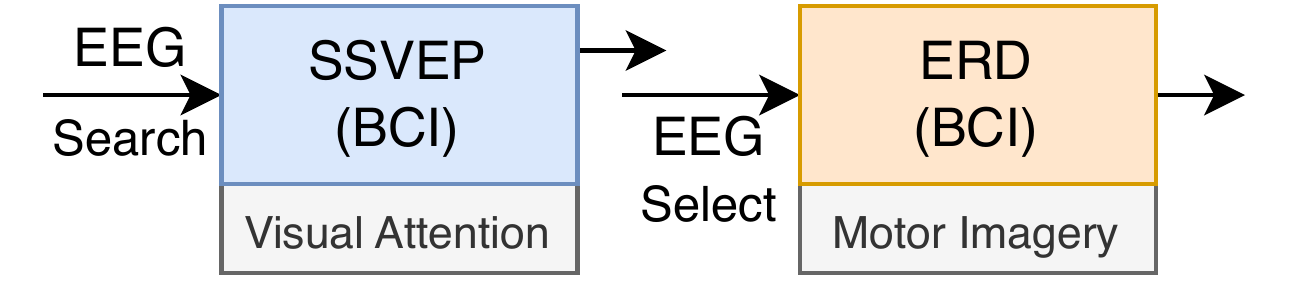}
        \caption{}
        \label{fig_hBCI_seq}
    \end{subfigure}
    \hfill
    \begin{subfigure}[b]{0.40\linewidth}
        \centering
        \includegraphics[width=\linewidth]{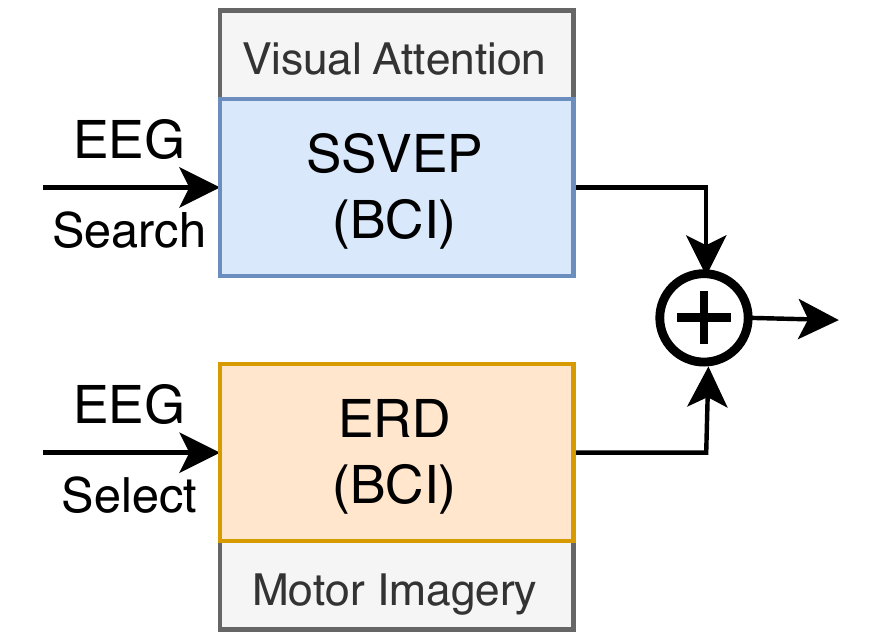}
        \caption{}
        \label{fig_hBCI_sim}
    \end{subfigure}
    \caption{Framework of hybrid BCI~\cite{pfurtscheller2010hybrid}: representing the data flow for (a) sequential and (b) simultaneous processing.}
    \label{fig_hBCI}
    \vspace{-3mm}
\end{figure}

A hBCI framework is implemented to overcome the inherent limitations of single-modality BCIs, namely the restricted command space of MI and the high cognitive fatigue or BCI-blindness with prolonged SSVEP use. A search-and-select paradigm designed to increase the number of discrete outputs is evaluated. Here, one modality serves as a high-level search/gating signal, while the other provides the final selection command. In the SSVEP-MI configuration, visual rhythmic entrainment (SSVEP) is used to navigate or search through interface layers, while motor intent executes the final selection. In MI-SSVEP, asynchronous motor intent serves as a trigger to activate the synchronous visual command space (SSVEP), effectively reducing false-positive activations during idle states.

Two operational modes were considered for this framework. In sequential mode (Fig~\ref{fig_hBCI_seq}), MI and SSVEP classifiers are applied to temporally distinct and non-overlapping trial windows. For example, a user first executes an MI trial and then attends to the SSVEP stimulus. In the simultaneous mode (Fig~\ref{fig_hBCI_sim}), the user co-activates both neural sources within the same window. Since our architecture uses a shared backbone for both, the error patterns remain consistent across modes. However, the hybrid approach offers a significant advantage over single modalities by creating $N_{MI} \times N_{SSVEP}$ total command outputs. While the classification accuracy is mathematically related, ITR varies significantly because simultaneous execution is faster than sequential gating.

\subsection{Quantisation-Aware Training}

To evaluate the deployability of each model under hardware resource constraints representative of embedded BCI devices, all five architectures were trained and evaluated across three numerical precisions: FP32, FP16, and INT8. Quantisation was implemented using PyTorch's native Torch Quantization framework with fake quantisation during the forward pass, enabling end-to-end QAT. Post-training quantisation (PTQ) was excluded because compact models with narrow weight distributions are especially vulnerable to INT8 accuracy collapse without in-training adaptation~\cite{jacob2018quantization}. For FP32, no quantisation was applied, whereas for FP16, native automatic mixed-precision was used and for INT8, the standard FBGEMM QAT configuration was applied. This staged design allows each model to learn to compensate for quantisation noise during training. SwitchBraidNet's architectural choices were specifically selected to mitigate failure modes of low-bit QAT, such as activation collapse and weight clustering. The hardsigmoid's bounded piecewise-linear form is empirically favoured for 8-bit QAT~\cite{bhalgat2020lsq}, to avoid the soft saturation of sigmoid that complicates scale estimation.

\subsection{Dataset and Preprocessing}

All experiments used the publicly available OpenBMI dataset~\cite{lee2019eeg}, comprising EEG recordings from 54 healthy subjects across two sessions, covering a two-class MI task (left vs. right hand) and a four-class SSVEP task (5.45, 6.67, 8.57, and 12 Hz). From the original 62-channel, 1000 Hz recordings, paradigm-specific electrode subsets were selected ($C$=20 motor-cortex channels for MI; 10 occipital channels for SSVEP) and downsampled to 256 Hz to balance computational efficiency with signal integrity. MI signals were bandpass-filtered at 8--30 Hz to isolate $\mu/\beta$ rhythms, and SSVEP signals at 4--40 Hz to preserve stimulus-driven harmonics. Trials were segmented into 1\,s windows with 0.5\,s overlap, yielding input of shape ($1 \times C \times 256$), providing sufficient frequency resolution for SSVEP decoding and temporal precision for event-related desynchronisation detection.

\subsection{Experimental Setup}

The models were implemented using PyTorch and MNE-Python, with training conducted over 50 epochs using the AdamW optimiser (LR=1e--3). The dataset was pre-divided into strictly disjoint training and testing sets following the OpenBMI protocol. The training set was further partitioned to form a validation set for model tuning. For low-bitwidth simulations, QAT was facilitated by the Torch Quantisation library, which inserted fake quantisation modules to model INT8 precision. All experiments were executed on an Nvidia RTX A2000 12GB GPU with CUDA 12.8, with model size estimation based on total parameter counts and their respective bit-widths. Five end-to-end deep learning architectures were evaluated, namely, DeepConvNet~\cite{schirrmeister2017deep}, EEGNet~\cite{lawhern2018eegnet}, ShallowConvNet~\cite{schirrmeister2017deep}, TSception~\cite{ding2022tsception} and the proposed SwitchBraidNet.

\subsection{Evaluation Metrics}

Three primary metrics (Accuracy, Cohen's Kappa, and ITR) were used to assess model performance across paradigms and quantisation levels. Classification accuracy was computed as the percentage of correctly classified trials over the held-out test set. To quantify classification performance while accounting for chance-level agreement, Cohen's Kappa ($\kappa$) was computed. To jointly quantify classification accuracy and command throughput, ITR~\cite{cecotti2024post, meena2015towards} was computed as follows. Let $P$ denote classification accuracy, $N$ the number of target classes, and $T$ the trial duration in seconds, then ITR is defined as~\cite{wolpaw2000brain}
\begin{equation}
\label{eq_itr_b}
    B = \log_{2}(N) + P\log_{2}(P) + (1-P)\log_{2}\left(\frac{1-P}{N-1}\right)
\end{equation}

where Eq.~\ref{eq_itr_b} gives information content per trial in bits, and ITR is calculated as $\text{ITR} = B \times (60/T)$ [bpm]. When $P<1/N$ is handled by clamping $B$ to zero to avoid undefined values. 

\input{tab_hbci}
\begin{table}[!t]
    \renewcommand{\arraystretch}{0.9}
    \centering
    \caption{Comparison of model complexity and memory footprints across varying numerical precisions}
    \label{tab_model_params}
    \setlength{\tabcolsep}{4pt}
    \begin{tabular}{lrrrr}
        \toprule
        & & \multicolumn{3}{c}{\textbf{Memory (KB)}} \\
        \textbf{Model} & \textbf{Parameters} & \textbf{32 Bit} & \textbf{16 Bit} & \textbf{8 Bit} \\
        \midrule
        DeepConvNet & 87,552 & 342.00 & 171.00 & 85.50 \\
        EEGNet & 2,292 & 8.95 & 4.48 & 2.24 \\
        ShallowConvNet & 17,962 & 70.16 & 35.08 & 17.54 \\
        TSception & 81,617 & 318.82 & 159.41 & 79.70 \\
        \textbf{SwitchBraidNet} & \textbf{3,106} & \textbf{12.13} & \textbf{6.07} & \textbf{3.03} \\
        \bottomrule
    \end{tabular}
\end{table}

Additionally, the effective model size was estimated using the relation $\text{Size (KB)} = [(\sum |\text{parameters}| \times 4) / 1024] \times (b/32)$, where the factor $b/32$ scales the standard FP32 footprint by the quantised bit-width $b \in \{32, 16, 8\}$. Statistical significance across models, operational modes, and numerical precisions was evaluated using the non-parametric paired Wilcoxon signed-rank test with Bonferroni correction, setting the significance threshold at $\alpha = 0.05$.


\section{Results}
\label{sec_results}

This section presents the classification performance of five deep learning models (DeepConvNet, EEGNet, ShallowConvNet, TSception, and SwitchBraidNet) across three BCI paradigms (MI, SSVEP, and hBCI) and three numerical precisions (FP32, FP16, and INT8). Single-modality and hBCI results are reported in Table~\ref{tab_hbci}, with model-parameter and memory statistics in Table~\ref{tab_model_params}.

\subsection{Single Modes}

\begin{figure}[!t]
    \centering
    \includegraphics[width=\linewidth]{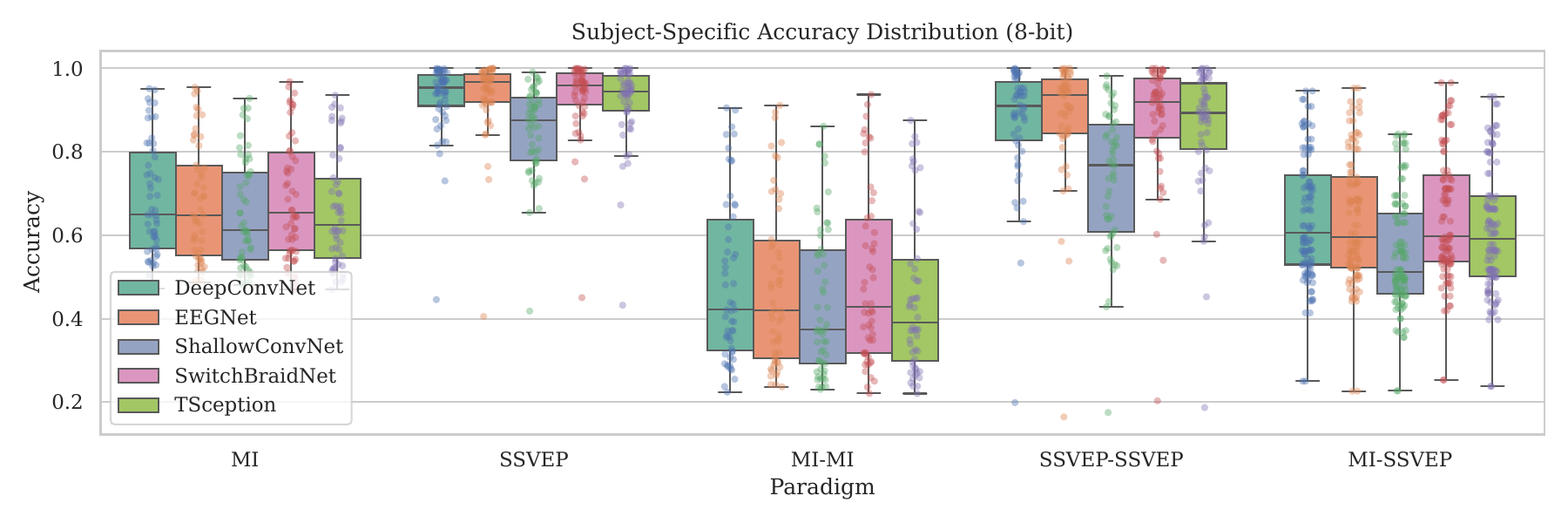}
    \caption{Distribution of subject-specific classification accuracies across 5-fold cross-validation}
    \label{fig_stability}
    \vspace{-3mm}
\end{figure}

On the binary MI task, classification accuracy ranged from 63.79\% to 69.49\% across models and precision levels. SwitchBraidNet achieved the highest accuracy at all three precisions, peaking at 69.49\% under FP16 ($\kappa = 0.39$). SwitchBraidNet achieved statistically significant performance improvements over EEGNet ($p < 0.001$) and DeepConvNet ($p < 0.001$). Overall model comparison between SwitchBraidNet and ShallowConvNet was non-significant ($p = 0.303$), confirming SwitchBraidNet's competitive baseline accuracy. SwitchBraidNet's ITR values represent a consistent improvement over EEGNet across all precisions. ShallowConvNet showed the largest drop under INT8, declining to 63.79\% ($-$4.50\% from FP32). On the four-class SSVEP task, all models achieved substantially higher accuracy than on MI. EEGNet attained the highest metrics across all precisions. SwitchBraidNet performed comparably to EEGNet, with a negligible gap of 0.17\%. DeepConvNet also performed at a level comparable to SwitchBraidNet. ShallowConvNet achieved the lowest SSVEP accuracy among the five models, with further degradation observed at INT8. Fig.~\ref{fig_stability} illustrates the subject-specific accuracy distribution across 5-fold cross-validation. SwitchBraidNet maintains a consistently elevated distribution and tighter spread across subjects.

\subsection{Hybrid Modes}

Sequential and simultaneous hBCI modes produced identical classification accuracy, F1-score, and $\kappa$ across all models and precision levels (p\,=\,1.0 for all three metrics), confirming that the two modes share the same decision quality. The simultaneous mode achieved a statistically significant higher ITR (59.0 vs.\ 29.5~bpm, $p < 0.001$), as co-acquiring signals within an overlapping window halves the effective trial duration from 2\,s to 1\,s (see Eq.~\ref{eq_itr_b}). Relative to the MI-alone baseline, the hybrid system achieved a statistically significant improvement in ITR ($+$23.8~bpm, p\,$<$\,0.001). Relative to the SSVEP-alone baseline, all hybrid metrics declined significantly (ITR: $-$57.9~bpm, p\,$<$\,0.001).

\subsection{Quantisation Robustness}

\begin{figure}[!t]
    \centering
    \includegraphics[width=\linewidth]{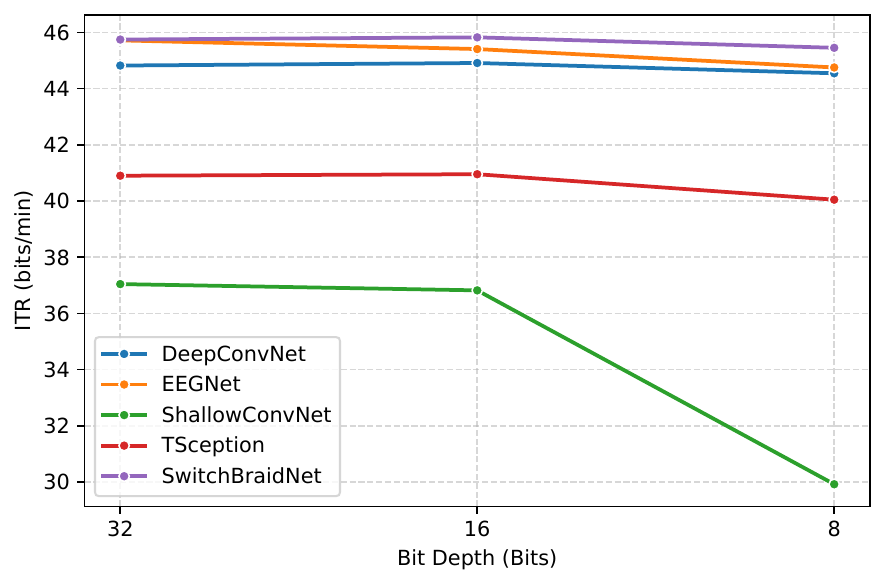}
    \caption{ITR vs.\ bit-depth across all architectures. SwitchBraidNet maintains the highest throughput and superior robustness at INT8 precision.}
    \label{fig_robust}
    \vspace{-3mm}
\end{figure}

Quantisation level had a statistically significant effect across all evaluated models ($p < 0.001$), with performance generally declining from 32-bit to 8-bit. The FP32-to-FP16 transition was effectively lossless for most models; for instance, SwitchBraidNet gained 0.16\% on MI and lost 0.06\% on SSVEP. Under INT8, SwitchBraidNet sustained the smallest performance degradation across both paradigms (0.21\% on MI, 0.08\% on SSVEP), $\approx$20$\times$ smaller than ShallowConvNet (4.50\% and 3.75\%, respectively). SwitchBraidNet achieved the best MI accuracy at INT8 (69.12\%) with a memory footprint of 3.03~KB; EEGNet achieved the best SSVEP accuracy at INT8 (93.67\%) at 2.24~KB. As shown in Fig.~\ref{fig_robust}, all architectures remain stable from FP32 to FP16 but diverge at INT8. ShallowConvNet's ITR drops from 37 to 30~bpm, while SwitchBraidNet maintains the highest baseline ITR ($\approx$46~bpm) with negligible degradation. EEGNet and TSception also exhibit high robustness to quantisation, though at lower ITR than SwitchBraidNet.


\section{Discussion}
\label{sec_discussion}

\subsection{Single-Mode Performance}

The narrow inter-model accuracy gap on MI ($<$6\%) reflects the inherently low spatial resolution of motor-imagery signals. SwitchBraidNet's marginal advantage stems from its dual-path temporal braid, which simultaneously captures $\mu/\beta$-band power envelopes across parallel branches without requiring explicit filter-bank segmentation. The LogVarLayer's mean-power readout further specialises the network towards band-power features, which are the physiologically motivated correlates of MI. ShallowConvNet also targets band-power but underperforms under INT8 quantisation, suggesting that its pooling window is more sensitive to precision-induced weight perturbations. This highlights a key trade-off: while heavier models or single paradigms (e.g., SSVEP) may yield higher peak accuracy, SwitchBraidNet prioritises robust hybrid performance and significantly reduced computational overhead for edge deployment.

The substantially higher SSVEP accuracies across all models reflect the high SNR of frequency-tagged responses. EEGNet's separable convolutional architecture compactly encodes the spectral-spatial structure of steady-state signals, which explains its lead on this task. SwitchBraidNet performed almost identically to EEGNet on SSVEP (0.17\% gap), indicating that its architecture generalises well beyond its primary MI design target. ShallowConvNet's lower SSVEP performance and INT8 sensitivity further suggest that its spatial pooling window lacks the granularity needed for multi-frequency discrimination.

\subsection{Hybrid vs. Single-Modality Trade-offs}

The hybrid paradigm introduces an accuracy cost relative to single-modality SSVEP while expanding the command space. Under FP32, the best observed SSVEP accuracy (93.65\%) far exceeds the best hybrid accuracy (64.81\%), which is expected: fusing the harder MI task ($\approx$69\%) with SSVEP across a higher-dimensional joint decision boundary inevitably increases the error rate.

Nevertheless, the hybrid paradigm offers practical advantages for real-world deployment. First, combining two-class MI with four-class SSVEP yields eight commands, enabling richer control without additional hardware. Second, periodically offloading to the MI channel reduces ocular strain from sustained SSVEP fixation~\cite{makri2015visual, diez2024assessment}. Third, grounding part of the command vocabulary in SSVEP improves reliability against MI non-stationarity~\cite{shenoy2006towards}. Fourth, the dual-modality design provides a functional fallback for users whose MI or SSVEP signal quality is situationally impaired.

The simultaneous hybrid ITR falls short of single-modality SSVEP because the accuracy reduction dominates the command-space expansion in the ITR formulation. For practical deployment, sequential mode may be preferred when simultaneous co-activation is cognitively prohibitive, whereas simultaneous mode remains attractive when maximising command bandwidth and user comfort. Translating high offline accuracies ($>0.9$) to practical online usage presents challenges, as unconstrained environments introduce noise, varying attention, and impedance degradation.

\subsection{Quantisation Robustness}

The lossless FP32-to-FP16 transition observed for most models is consistent with the known regularising effect of reduced-precision arithmetic, which introduces small stochastic perturbations during training analogous to mild noise injection~\cite{jacob2018quantization}. This is consistent with the marginal accuracy gain of SwitchBraidNet at FP16 on MI (0.16\%).

Under INT8 - the most practically relevant precision for microcontroller and FPGA deployment (e.g., ARM Cortex-M) - SwitchBraidNet's $\approx$20$\times$ smaller degradation relative to ShallowConvNet indicates that its temporal pathways and spatial switching preserve the features required for MI and SSVEP decoding under constrained numerical precision. SSVEP models' overall robustness to quantisation (p\,=\,0.074) further suggests that frequency-domain features are less sensitive to weight precision than the spatial features underpinning MI decoding. For SSVEP-only deployment, EEGNet ($2.24$~KB at INT8) represents the optimal choice by accuracy. For hybrid deployment, SwitchBraidNet's consistent top-2 performance across both modalities and its small footprint make it the most versatile model.

\subsection{Limitations}
While robust to quantisation, SwitchBraidNet faces limitations. In noisy environments or for subjects with subtle neural features, INT8 precision may cause feature collapse and degrade performance. Additionally, aggressive quantisation may reduce the fidelity of high-frequency bands, requiring band-specific precision scaling. Finally, validating on additional datasets (e.g., BCI IV 2a/2b) remains a priority to confirm generalizability beyond the OpenBMI dataset.

\subsection{ITR Threshold Analysis}

\begin{figure}[!t]
    \centering
    \includegraphics[width=\linewidth]{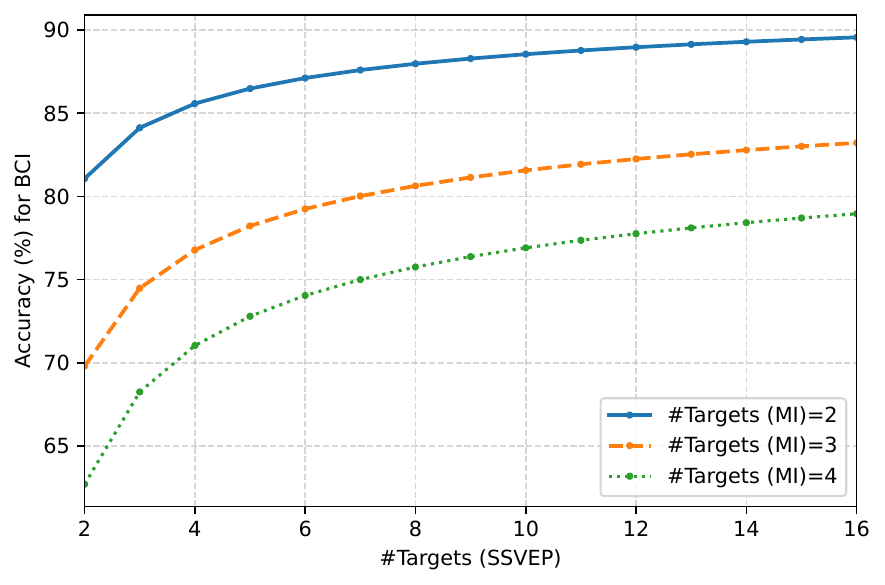}
    \caption{Minimum MI accuracy required for hBCI to exceed pure SSVEP ITR versus SSVEP target count $N_\text{SSVEP}$ (assuming ideal SSVEP decoding).}
    \label{fig_itr}
    \vspace{-3mm}
\end{figure}

Fig.~\ref{fig_itr} maps the minimum MI accuracy required for the hBCI ITR to exceed that of a pure SSVEP system, for $N_\text{MI} \in \{2,3,4\}$ and assuming perfect SSVEP accuracy as a theoretical upper bound. For $N_\text{MI}=2$, the required minimum at $N_\text{SSVEP}=2$ is 81.1\%, rising and asymptoting as $N_\text{SSVEP}$ increases well above the $\approx$69\% MI accuracy observed across all five architectures. This finding is consistent with O'Doherty et al.~\cite{o2014exploring}, who reported an analogous 86\% threshold for a gaze-MI hybrid. Moreover, this bound is strict; at empirically observed SSVEP accuracies ($\approx$93.5\%), the SSVEP baseline ITR decreases, further narrowing the gap. Hybrid deployment is therefore justified primarily by complementary advantages: expanded command vocabulary, reduced visual fatigue, and fallback resilience, rather than by net ITR gain over SSVEP alone.


\section{Conclusion}
\label{sec_conclusion}

This paper presented a systematic evaluation of five deep learning architectures for hybrid EEG-based BCIs under QAT, utilising the OpenBMI dataset across MI, SSVEP, and hBCI paradigms. The central contribution is SwitchBraidNet, a lightweight three-stage architecture comprising a dual-path temporal braid, a squeeze-and-excitation spatial switch, and a log-variance layer. It achieves the highest MI and hBCI classification accuracy across all tested precisions (INT8 accuracy: 69.12\% MI, 64.81\% hBCI; footprint: 3.03~KB), with negligible INT8 degradation (0.21\%) and ITRs of 32.41 and 64.82~bpm in sequential and simultaneous modes, respectively. On SSVEP, EEGNet leads marginally, but SwitchBraidNet remains statistically competitive, with both exhibiting negligible INT8 degradation. The QAT benchmark and ITR framework presented here provide a reproducible foundation for future research in neural signal processing and hardware-aware deep learning, with a natural extension being the incorporation of a third modality, such as P300, to further expand the command space. Future work will explore optimising the fusion strategy using adaptive weighting and attention mechanisms to further boost hybrid performance.


\begin{acknowledgements}
This study was supported by the IP/IITGN/CSE/YM/2324/05 grant.
\end{acknowledgements}


\section*{Bibliography}
\bibliography{references}


\end{document}

%% file: tab_hbci.tex
\begin{table*}
  \centering
  \caption{Performance metrics for single modality paradigms and hybrid BCI systems in sequential and simultaneous modes}
  \label{tab_hbci}
  \setlength{\tabcolsep}{4pt}
  \resizebox{\textwidth}{!}{
  \begin{tabular}{llccccccccccccccc}
  \toprule
     &  & \multicolumn{3}{c}{DeepConvNet} & \multicolumn{3}{c}{EEGNet} & \multicolumn{3}{c}{ShallowConvNet} & \multicolumn{3}{c}{TSception} & \multicolumn{3}{c}{SwitchBraidNet} \\
    \cmidrule(lr){3-5} \cmidrule(lr){6-8} \cmidrule(lr){9-11} \cmidrule(lr){12-14} \cmidrule(lr){15-17}
    \textbf{Paradigm} &  & 32 & 16 & 8 & 32 & 16 & 8 & 32 & 16 & 8 & 32 & 16 & 8 & 32 & 16 & 8 \\
    \midrule
    \multirow{4}{*}{MI} & Acc & 68.57 & 68.75 & 68.39 & 69.08 & 68.67 & 68.04 & 68.29 & 68.58 & 63.79 & 65.73 & 66.09 & 65.48 & \textbf{69.33} & \textbf{69.49} & \textbf{69.12} \\
     & F1 & 67.91 & 68.44 & 68.35 & 69.06 & 68.46 & 68.02 & 68.08 & 68.58 & 62.50 & 65.65 & 66.05 & 65.47 & \textbf{69.23} & \textbf{69.27} & \textbf{68.92} \\
     & ITR & 6.12 & 6.24 & 5.99 & 6.46 & 6.18 & 5.76 & 5.93 & 6.12 & 3.33 & 4.36 & 4.56 & 4.22 & \textbf{6.64} & \textbf{6.75} & \textbf{6.50} \\
     & Kappa & 0.37 & 0.38 & 0.37 & 0.38 & 0.37 & 0.36 & 0.37 & 0.37 & 0.28 & 0.31 & 0.32 & 0.31 & \textbf{0.39} & \textbf{0.39} & \textbf{0.38} \\
    \midrule
    \multirow{4}{*}{SSVEP} & Acc & 93.32 & 93.26 & 93.24 & \textbf{93.65} & \textbf{93.71} & \textbf{93.67} & 86.77 & 86.33 & 83.02 & 92.21 & 91.99 & 91.66 & 93.48 & 93.42 & 93.40 \\
     & F1 & 93.33 & 93.29 & 93.25 & \textbf{93.67} & \textbf{93.73} & \textbf{93.69} & 86.78 & 86.34 & 82.80 & 92.24 & 92.01 & 91.66 & 93.49 & 93.44 & 93.41 \\
     & ITR & 92.42 & 92.21 & 92.15 & \textbf{93.49} & \textbf{93.70} & \textbf{93.57} & 73.58 & 72.46 & 64.42 & 88.90 & 88.22 & 87.22 & 92.94 & 92.74 & 92.69 \\
     & Kappa & 0.91 & 0.91 & 0.91 & \textbf{0.92} & \textbf{0.92} & \textbf{0.92} & 0.82 & 0.82 & 0.77 & 0.90 & 0.89 & 0.89 & 0.91 & 0.91 & 0.91 \\

    \midrule
    
     & Acc & 63.99 & 64.12 & 63.77 & 64.69 & 64.35 & 63.73 & 59.25 & 59.20 & 52.96 & 60.61 & 60.79 & 60.01 & \textbf{64.81} & \textbf{64.92} & \textbf{64.56} \\
    Hybrid & F1 & 63.38 & 63.84 & 63.74 & 64.68 & 64.17 & 63.73 & 59.08 & 59.21 & 51.76 & 60.56 & 60.77 & 60.01 & \textbf{64.73} & \textbf{64.73} & \textbf{64.38} \\
    MI-SSVEP / & Kappa & 0.59 & 0.59 & 0.59 & \textbf{0.60} & 0.59 & 0.59 & 0.53 & 0.53 & 0.46 & 0.55 & 0.55 & 0.54 & \textbf{0.60} & \textbf{0.60} & \textbf{0.60} \\
    SSVEP-MI & ITR (Seq) & 31.39 & 31.53 & 31.14 & 32.16 & 31.79 & 31.11 & 26.42 & 26.38 & 20.46 & 27.81 & 28.00 & 27.20 & \textbf{32.29} & \textbf{32.41} & \textbf{32.02} \\
     & ITR (Sim) & 62.78 & 63.05 & 62.29 & 64.32 & 63.58 & 62.22 & 52.85 & 52.75 & 40.91 & 55.62 & 55.99 & 54.40 & \textbf{64.58} & \textbf{64.82} & \textbf{64.04} \\
    \bottomrule
  \end{tabular}}
\end{table*}